\documentclass[11pt, A4paper]{article}
	\usepackage{amsmath}
	\usepackage{amssymb}
	\usepackage{mathrsfs}
	\usepackage{graphicx}
	\usepackage{cmbright}
	\usepackage[symbol]{footmisc}

	\def\aa{A\&A}

	\def\dommed{\mathscr M}
	\def\permit{\epsilon}
	\def\permea{\mu}
	\def\indr{n}
	\def\ham{\mathcal H}
	\def\metr{g}
	\def\bmetr{\bar g}
	\def\ee{\mathrm{e}}
	\def\ii{\mathrm{i}}
    \def\dd{\mathrm{d}}
	\def\eik{\mathscr{S}}
	\def\obs{\mathcal{O}}
	\def\cone{\mathscr{C}}
	\def\wrdline{\mathscr{L}}
	\def\ttf{\mathcal T}
	\def\domint{\mathscr{D}}

	\newcommand{\Qvect}[1]{\boldsymbol{#1}}
	\newcommand{\vect}[1]{\vec{#1}}
	\newcommand{\cvect}[2]{{#1}{}^{#2}}
	\newcommand{\cflin}[2]{{#1}{}_{#2}}
    
	\newcommand{\Poiss}[2]{\left\{#1,#2\right\}}
	
	\newcommand{\lightray}[2]{\mathscr{G}_{#1}^{[#2]}}

\begin{document}
	
	\setcounter{figure}{0}
	\setcounter{table}{0}
	\setcounter{footnote}{0}
	\setcounter{equation}{0}
	
	\noindent {\Large\bf TIME AND FREQUENCY TRANSFERS IN OPTICAL SPACETIME}
	\vspace*{0.7cm}

	\noindent\hspace*{1cm} A. BOURGOIN$^{1,}$\footnote[2]{adrien.bourgoin\,AT\,obspm.fr}, P. TEYSSANDIER$^1$, P. TORTORA$^2$, and M. ZANNONI$^2$ \\[0.2cm]
	\noindent\hspace*{1cm} $^1$ SYRTE, Observatoire de Paris, PSL Research University, CNRS, Sorbonne Université, UPMC, Univ. Paris 6, 61 avenue de l'Observatoire, 75014 Paris - France\\[0.1cm]
	\noindent\hspace*{1cm} $^2$ Dipartimento di Ingegneria Industriale, Alma Mater Studiorum - Università di Bologna, Via Fontanelle 40, 47121 Forlì - Italy\\
	
	\vspace*{1cm}
	
	\noindent {\large\bf ABSTRACT.} Solving the null geodesic equations for a ray of light is a difficult task even considering a stationary spacetime. The problem becomes even more difficult if the electromagnetic signal propagates through a flowing optical medium. Indeed, because of the interaction between light and matter, the signal does not follow a null geodesic path of the spacetime metric anymore. However, having a clear description of how the time and frequency transfers are affected in this very situation is of a prime importance in astronomy. As a matter of fact, ranging to satellites and Moon, very long baseline interferometry, global navigation satellite systems, and radio occultation experiments are few examples of techniques involving light propagation in flowing optical media. By applying the time transfer functions formalism to optical spacetime, we show that the time and frequency transfers can be determined iteratively up to any desired order within the quasi-Minkowskian path approximation. We present some applications in the context of radio occultation experiments and discuss possible future applications to the modeling of tropospheric delays. 
	
	\vspace*{1cm}
	
	\noindent {\large\bf 1. INTRODUCTION}
	\smallskip
	
\begin{figure}[t]
	  \begin{center}
	    \includegraphics[scale=0.362]{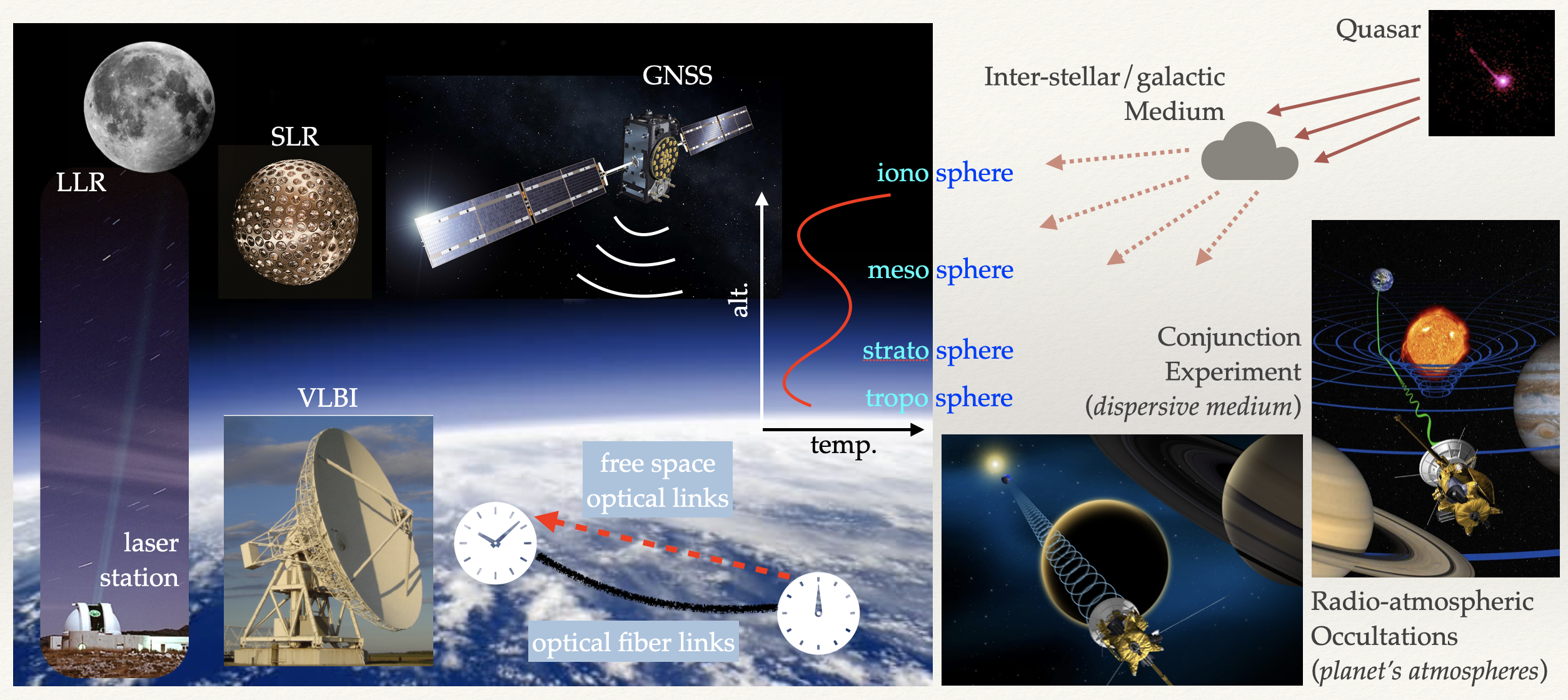}
	    \caption{Illustration of different techniques in astronomy and in metrology requiring an accurate modeling of light propagation in optical media. For astro-geodetic techniques (i.e., SLR, LLR, GNSS, VLBI, etc.) the main optical medium is the Earth's atmosphere. In radio science, during an occultation event the optical medium is the atmosphere of a distant planet or a Moon (e.g., Titan). In metrology, when performing clock comparison, the main optical medium is either the Earth's atmosphere for free space optical links or a fiber when performing optical fiber links.}
		\label{fig_01}
      \end{center}
	\end{figure}

	Light propagation in optical media is everywhere in astronomy and metrology (see few examples in Fig. \ref{fig_01}). Given the precision of some techniques---at the level of mm for satellite laser ranging (SLR), cm for lunar laser ranging (LLR), $\mu$as for very long baseline interferometry (VLBI), tenth of cm for the range and $\mu$m/s for the range-rate of current space probes (e.g., BepiColombo, JUICE, etc.)---the modeling of the time and frequency transfer requires one to properly describe the light propagation in a vacuum and in optical media too. If light propagates along null geodesics of spacetime in a vacuum, it does not in an optical medium due to the electromagnetic interaction. However, for linear non-dispersive media, refractivity acts on optical rays as gravity would on test masses, and hence, in some circumstances, refractivity can be interpreted as curvature within the formalism of the optical spacetime metric (also known as Gordon's metric). Then, when ligh propagates  into such non-dispersive media, optical rays actually follow null geodesics of the optical spacetime. Solving null geodesic equations is a difficult task which can nevertheless be tackled efficiently within the formalism of time transfer functions (see Le Poncin-Lafitte \emph{et al.}, 2004). In this work, we make use of the time transfer functions formalism considering an optical metric in order to accurately model the time and frequency transfers in every situations involving light propagation in a linear non-dispersive medium. The results that are derived are applied in the context of radio occultations experiments and then compared to numerical integration of the equations of relativistic geometrical optics. The light-dragging effect is shown to be at the threshold of visibility in many experiments and hence shall be carefully modeled in the next future. The formalism exposed here, and discussed in more depth in Bourgoin 2020 and Bourgoin \emph{et al.}, 2021, offers an elegant framework to deal with light propagation in flowing media.

	\newpage
    \vspace*{-0.82cm}
	{\large\bf 2. NOTATIONS AND CONVENTIONS}
	\smallskip

	In this paper, $c$ denotes the speed of light in a vacuum.
	
	Spacetime is assumed to be covered by a global quasi-Cartesian coordinate system $( \cvect{x}{0} , \cvect{x}{1} , \cvect{x}{2}$, $\cvect{x}{3} )$, where $\cvect{x}{0} = ct$, $t$ having the dimension of a time and $\cvect{x}{1}$, $\cvect{x}{2}$, and $\cvect{x}{3}$ the dimension of a length.
	
	Point-events in spacetime are denoted by capital letters such as $A$, $B$, etc. We use Greek letters for indices running from 0 to 3, and Latin letters for indices running from 1 to 3. Any bold letter refers to an ordered quadruple of numbers, namely $\Qvect{x} = ( \cvect{x}{0} , \cvect{x}{1} , \cvect{x}{2} , \cvect{x}{3} ) = ( \cvect{x}{\alpha} )$. We use a vector-like notation for any ordered triple, that is to say $\vect{a} = ( \cvect{a}{1} , \cvect{a}{2} , \cvect{a}{3} ) = ( \cvect{a}{i} )$. So, the coordinates of a point-event $P$ in spacetime will be denoted by $\Qvect{x}_P = ( ct_P , \vec{x}_P )$. 
	
	According to Einstein's convention on repeated indices, expressions like $\cvect{x}{\mu} \cflin{y}{\mu}$ and $\cvect{a}{i} \cflin{b}{i}$ stand for $\sum_{\mu=0}^{3} \cvect{x}{\mu} \cflin{y}{\mu}$ and $\sum_{i=1}^{3} \cvect{a}{i} \cflin{b}{i}$, respectively.
	
	Throughout this work, the Lorentzian metric describing the gravitational field and determining the proper time of the observers is denoted by $\metr$ and called the \emph{physical metric}. The signature adopted for $\metr$ is $(+,-,-,-)$.
	
	The scalar product between 4-vectors $\Qvect{x}$ and $\Qvect{y}$ is denoted by $\Qvect{x} \cdot \Qvect{y} = \metr_{\mu\nu} \cvect{x}{\mu} \cvect{y}{\nu}$. The scalar product between 3-vectors $\vect{a}$ and $\vect{b}$ in the Euclidean 3-space is denoted by $\vect{a} \cdot \vect{b} = \delta_{ij} \cvect{a}{i} \cvect{b}{j}$, where $\delta_{ij}$ is the Kronecker symbol. $\vec{a}\times\vec{b}$ is the triple obtained by the usual rule giving the exterior product of two vectors of the ordinary Euclidean 3-space.
	
	The partial derivative $\partial f/\partial x^{\mu}$ is often denoted by $f_{,\mu}$ for the sake of brevity. A quantity $f$ when evaluated at point-event $P$ is denoted either by $f(P)$ or by $(f)_P$.

	\vspace*{0.7cm}
	\noindent {\large\bf 3. RELATIVISTIC GEOMETRICAL OPTICS}
	\smallskip

    In this section, we recall some fundamental results of geometrical optics in the context of metric theories of gravity (see e.g., Synge 1960). We start by commenting on the geometrical optics approximation, and then provide expressions for the usual quantities that an observer can measure in its local rest frame, such as the frequency.

    \newpage

	\noindent {\large\bf 3.1 Physical properties of the optical medium}
	\smallskip

	We assume that a bounded domain $\dommed$ of spacetime is filled with a medium that is \emph{linear}, \emph{isotropic}, and \emph{non-dispersive}. Neat vacuum is assumed outside $\dommed$. The optical medium inside $\dommed$ is characterized by two scalar functions: the \emph{permittivity} $\permit(P)$ and the \emph{permeability} $\permea(P)$, where $P \in \dommed$. Outside $\dommed$ these functions reduce to their vacuum values, namely $\permit_0$ and $\permea_0$, respectively. The speed of light in a vacuum is given by $c = ( \permit_0 \permea_0 )^{-1/2}$. The index of refraction $\indr$ of the medium is a scalar field defined by
	\begin{equation}
	  \indr (P) = c \sqrt{ \permit (P) \permea (P) } .
	\end{equation}
    From these definitions, if $P \notin \dommed$, then $\indr (P) = 1$ as it might be expected in a vacuum.

	We also assume that the fluid elements are not colliding, which means that their worldlines form a congruence of time-like curves. Accordingly, observers for which a fluid element is momentarily at rest is a \emph{comoving observer with the medium}. The components of the unit 4-velocity vector of the fluid element heading through $P$ is denoted by $\cvect{w}{\mu}(P)$ or simply $\cvect{w}{\mu}$, when there exists no ambiguity. $\cvect{w}{\mu}(P)$ is clearly a vector field defined in $\dommed$.

	\vspace*{0.3cm}
	\noindent {\large\bf 3.2 Light propagating through the optical medium}
	\smallskip

    Henceforth, we work within the geometrical optics approximation assuming an optical medium whose properties are summarized in Sect. 3.1. We assume that spacetime is swept by an electromagnetic wave described by a set of functions of the type:
	\begin{equation}
	  \psi (P) = \mathscr A (P) \, \ee^{\ii \eik (P)} .
	\end{equation}
	where $\mathscr A(P)$ is the \emph{amplitude} varying slowly at the scale of the wavelength of the signal and $\eik (P)$ is the \emph{phase function} (or an \emph{eikonal function}) changing rapidly between $-\infty$ and $+\infty$. The wavelength is assumed to be way smaller than spacetime curvature.
 
	It can be inferred (Synge 1960; Bourgoin 2020) from Maxwell's equations treated within the geometrical optics approximation, that any light ray passing through $\dommed$ has the following properties:

	\begin{enumerate}
 	  \item The phase $\eik$ satisfies an \emph{eikonal equation}
	  \begin{equation}
		\left( \bmetr^{\mu\nu} \, \partial_\mu \eik \, \partial_\nu \eik \right)_P = 0 ,
		\label{eq:eik}
	  \end{equation}
	  at any point-event $P$ along the ray; the quantities $\bmetr^{\mu\nu}$ being given by
	  \begin{equation}
		\bmetr^{\mu\nu} = \metr^{\mu\nu} + \left( n^2 - 1 \right) w^\mu w^\nu .
	  \end{equation}

	  \item The ray is a null geodesic of the \emph{optical metric} $\bmetr_{\mu\nu}$ (also called Gordon's metric) defined by 
	  \begin{equation}
		  \bmetr_{\mu\nu} = \metr_{\mu\nu} - \left( 1 - \frac{1}{n^2} \right) \cflin{w}{\mu} \cflin{w}{\nu} .
	  \end{equation}
	  It is important to note that the quantities $\bmetr_{\mu\nu}$ are such that $\bmetr^{\mu\alpha} \bmetr_{\alpha \nu} = \delta^{\mu}{}_{\nu}$.

	  \item For a suitable choice of the affine parameter $\lambda$ along the trajectory of the ray $\cvect{x}{\mu} = \cvect{x}{\mu} (\lambda)$, the vector tangent to the ray is the \emph{optical wave-vector} $\bar{\Qvect{k}}$ whose components are given by
	  \begin{equation}
		\cvect{\bar{k}}{\mu} = \bmetr^{\mu\nu} \cflin{k}{\nu} ,
		\label{eq:bkvec}
	  \end{equation}
	  $\cflin{k}{\nu}$ being the components of the fundamental \emph{wave-covector} defined as
	  \begin{equation}
		\cflin{k}{\nu} \equiv \partial_\nu \eik . \label{eq:flink}
	  \end{equation}

	  \item The differential equations describing the trajectory of the ray can be derived from the following \emph{Hamilton function} (corresponding to the eikonal Eq. \eqref{eq:eik}):
	  \begin{equation}
	    \ham ( \cvect{x}{\mu} , \cflin{k}{\nu} ) = \frac{1}{2} \, \bmetr^{\mu\nu} \cflin{k}{\mu} \cflin{k}{\nu} ,
	  \end{equation}
	  where $\cvect{x}{\mu}$ and $\cflin{k}{\nu}$ are regarded as a set of \emph{independent} and \emph{canonical variables}. The propagation rays are thus the curves $\cvect{x}{\mu} = \cvect{x}{\mu} (\lambda)$ being solutions to the canonical equations
	  \begin{equation}
		\frac{\dd \cvect{x}{\mu}}{\dd \lambda} = \Poiss{\ham}{\cvect{x}{\mu}} , \qquad \frac{\dd \cflin{k}{\mu}}{\dd \lambda} = \Poiss{\ham}{\cflin{k}{\mu}} ,
	  \end{equation}
	  where $\Poiss{f}{g}$ denotes the Poisson's bracket of $f = f( \cvect{x}{\mu} , \cflin{k}{\nu} )$ and $g = g( \cvect{x}{\mu} , \cflin{k}{\nu} )$, namely
	  \begin{equation}
		\Poiss{f}{g} = \frac{\partial f}{\partial \cflin{k}{\nu}} \frac{\partial g}{\partial \cvect{x}{\nu}} - \frac{\partial f}{\partial \cvect{x}{\nu}} \frac{\partial g}{\partial \cflin{k}{\nu}} .
	  \end{equation}
	  The ray-tracing equations thus read as:
	  \begin{subequations}\label{eq:caneqs}
		\begin{align}
		  \frac{\dd \cvect{x}{\mu}}{\dd \lambda} & = \cvect{\bar k}{\mu} = \cvect{k}{\mu} + (n^2-1) (\Qvect{k} \cdot \Qvect{w}) \cvect{w}{\mu}, \\
		  \frac{\dd \cflin{k}{\alpha}}{\dd \lambda} & = - \frac{1}{2} \, \metr^{\mu\nu}{}_{\!,\alpha} \cflin{k}{\mu} \cflin{k}{\nu} - (n^2 - 1) (\Qvect{k} \cdot \Qvect{w}) \cflin{k}{\nu} \cvect{w}{\nu}{}_{\!,\alpha} - n n_{,\alpha} (\Qvect{k} \cdot \Qvect{w})^2 . \label{eq:caneqsb}
		\end{align}
	  \end{subequations}
	  where $\cvect{k}{\mu}$ are the components of the \emph{wave-vector} $\Qvect{k}$, namely:
	  \begin{equation}
		\cvect{k}{\mu} = \metr^{\mu\nu} \cflin{k}{\nu} . 
		\label{eq:kvec}
	  \end{equation}
	  Let us emphasize that the components of the wave-covector are raised making use of the physical metric in Eq. \eqref{eq:kvec} whereas they are raised using the optical metric in Eq. \eqref{eq:bkvec}.
		  
	  \item The Hamilton function $\ham ( \cvect{x}{\mu} , \cflin{k}{\nu} )$ and phase $\eik (\cvect{x}{\mu})$ are \emph{first integrals} during propagation of the ray, since
	  \begin{equation}
		\frac{\dd \ham}{\dd \lambda} = \Poiss{\ham}{\ham} = 0 , \qquad \frac{\dd \eik}{\dd \lambda} = \Poiss{\ham}{\eik} = \bmetr^{\mu\nu} \cflin{k}{\mu} \cflin{k}{\nu} = 0 .
		\label{eq.firstints}
	  \end{equation}
	  The last equality of the second equation results from the eikonal equation \eqref{eq:eik} and the definition \eqref{eq:flink}. The constancy of $\eik$ is at the core of the \emph{time transfer functions} formalism discussed in Sect. 4.
	\end{enumerate}

	\vspace*{0.3cm}
	\noindent {\large\bf 3.3 Doppler effect}
	\smallskip

	Let us consider a particle (a photon) with 4-momentum $\Qvect{p}$ and an observer $\obs$ of unit 4-velocity $\Qvect{u} (\tau)$ at instant $\tau$ of $\obs$'s proper time. It is well known that the energy $E$ of the particle measured by $\obs$ at $\tau$ is given by $E = c \Qvect{p} \cdot \Qvect{u}$ (see e.g., Synge 1960). Then, using the well-known de Broglie's wave-particle duality relations: $\Qvect{p} = \hbar \Qvect{k}$ with $\hbar$ the reduced Planck constant, we infer that the frequency $\nu$ of the particle measured by $\obs$ at instant $\tau$ is given by the following expression:
	\begin{equation}
	  \nu = \frac{c}{2\pi} \Qvect{k} \cdot \Qvect{u} .
	  \label{eq:nufreq}
	\end{equation}

\begin{figure}[t]
	  \begin{center}
	    \includegraphics[scale=0.5]{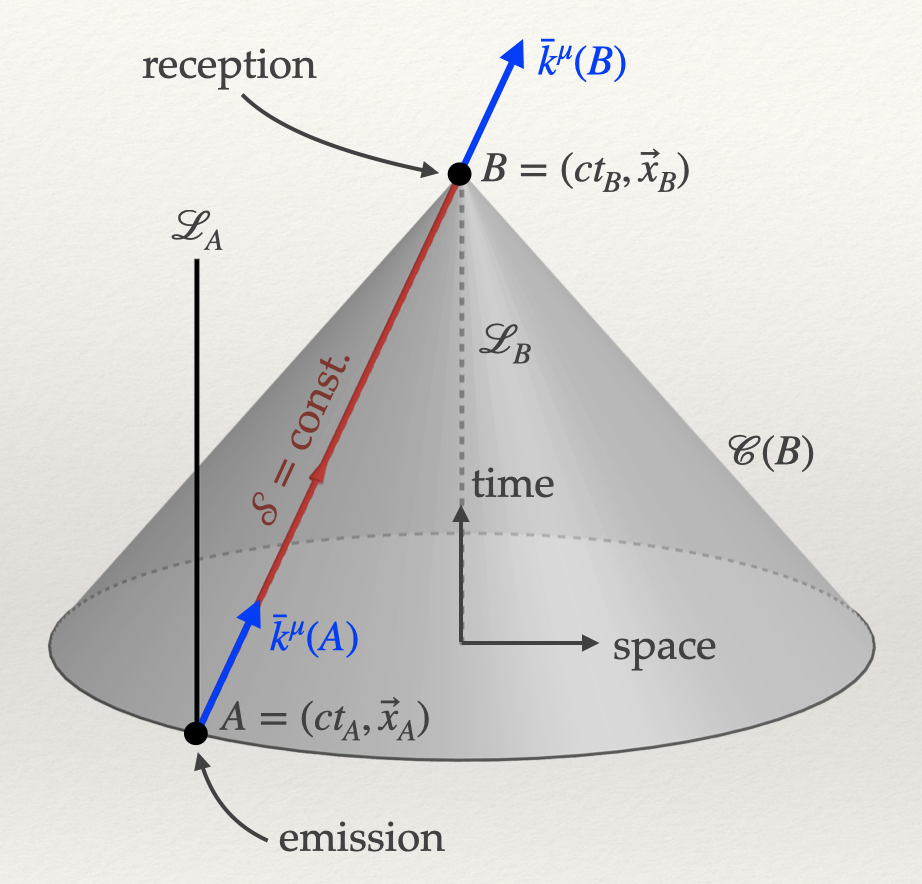}
	    \caption{Spacetime diagram representing a time transfer between an emitter $\obs_A$, at point-event $A$, and a receiver $\obs_B$, at point-event $B$. The signal (\emph{in red}) propagates along $\cone(B)$, the past-light cone at $B$. The worldlines of observers $\obs_A$ and $\obs_B$ are denoted $\wrdline_A$ and $\wrdline_B$, respectively.}
	    \label{fig_02}
	  \end{center}
	\end{figure}

	Let us introduce $\obs_A$ and $\obs_B$, two observers of unit 4-velocities $\Qvect{u}_A (\tau_A)$ and $\Qvect{u}_B (\tau_B)$ with proper times $\tau_A$ and $\tau_B$, respectively. Let suppose that $\obs_A$ and $\obs_B$ intersect the worldine of the particle at two different point-events $A$ and $B$, respectively (see Fig. \ref{fig_02}). By convention, we consider that $\obs_A$ is located in the past of $\obs_B$, so we call $\obs_A$ the \emph{emitter} and $\obs_B$ the \emph{receiver}. According to Eq.~\eqref{eq:nufreq}, the frequencies $\nu_B$ and $\nu_A$ measured by observers $\obs_B$ and $\obs_A$ are thus related by
	\begin{equation}
	  \frac{\nu_B (t_B)}{\nu_A (t_A)} = \frac{ \Qvect{k} (B) \cdot \Qvect{u}_B (\tau_B (t_B)) }{ \Qvect{k} (A) \cdot \Qvect{u}_A (\tau_A (t_A))} ,
	  \label{eq:nuBsnuA}
	\end{equation}
	where the different quantities are given here as functions of the (global) coordinate time $t$.

	\vspace*{0.7cm}
	\noindent {\large\bf 4. TIME TRANSFER FUNCTIONS FORMALISM}
	\smallskip

	In this section, we recall how the \emph{time transfer} and the \emph{frequency transfer} can be determined from \emph{time transfer functions}. Then, we also present the general method to determine the time transfer function at any order within the quasi-Minkowskian regime.

	\vspace*{0.3cm}
	\noindent {\large\bf 4.1 The time transfer function}
	\smallskip

	The constancy of $\eik$ during the ray propagation (cf. Eq. \eqref{eq.firstints}) implies that the eikonal function at emission $A$ and reception $B$ satisfies
	\begin{equation}
	  \eik (ct_B,\vect{x}_B) = \eik ( ct_A,\vect{x}_A ) .
	  \label{eq:eikconst}
	\end{equation}
	Therefore, it is natural to express, for instance, $t_A$, the coordinate time at emission, as a function of $t_B$, $\vect{x}_A$, and $\vect{x}_B$ (see Fig. \ref{fig_02}). Therefore, the (coordinate) \emph{time transfer} is expressed such as
	\begin{equation}
	  t_B - t_A = \ttf ( \vec{x}_A , t_B , \vec{x}_B ) ,
	  \label{eq:defttf}
	\end{equation}
	where $\ttf ( \vec{x}_A , t_B , \vec{x}_B )$ is the \emph{time transfer function} at reception---a time transfer function at emission also exists; it depends on the time of emission and the positions of both the emitter and receiver.
	
	After having substituted for $t_A$ from Eq. \eqref{eq:defttf} into \eqref{eq:eikconst} and after having differentiated the expression with respect to $\cvect{x}{i}_A$, $t_B$, and $\cvect{x}{i}_B$, we find the following relationships
	\begin{subequations}\label{eq:dirstat}
	  \begin{align}
		\left(\cflin{l}{i}\right)_A & \equiv \left(\frac{\cflin{k}{i}}{\cflin{k}{0}}\right)_{\!A} = c \frac{\partial \ttf}{\partial \cvect{x}{i}_A} , \\
		\left(\cflin{l}{i}\right)_B & \equiv \left(\frac{\cflin{k}{i}}{\cflin{k}{0}}\right)_{\!B} = - c \frac{\partial \ttf}{\partial \cvect{x}{i}_B} \left(1 - \frac{\partial \ttf}{\partial t_B} \right)^{-1} , \\
		(\cflin{k}{0})_B & \Bigg. = (\cflin{k}{0})_A \left( 1 - \frac{\partial \ttf}{\partial t_B} \right) . \label{eq:dirstatc}
	  \end{align}	
	\end{subequations}
	
	The \emph{frequency transfer} can thus be expressed in term of the time transfer function. After inserting equations \eqref{eq:dirstat} into \eqref{eq:nuBsnuA}, the Doppler frequency change reads as follows
	\begin{equation}
	  \frac{\nu_B}{\nu_A} = \frac{( \cvect{u}{0} )_B}{( \cvect{u}{0} )_A} \, \frac{(\cflin{k}{0})_B}{(\cflin{k}{0})_A} \, \frac{\left( 1 + \cvect{\beta}{i} \cflin{l}{i} \right)_B}{\left( 1 + \cvect{\beta}{i} \cflin{l}{i} \right)_A} ,
	  \label{eq:nuBsnuAli}
	\end{equation}
	where $(\cvect{u}{0})_A$ and $(\cvect{u}{0})_B$ are given by
	\begin{subequations}
	  \begin{align}
	    (\cvect{u}{0})_A & = \sqrt{\metr_{00} (A) + 2 \metr_{0i} (A) \, \cvect{\beta}{i}_A + \metr_{ij} (A) \, \cvect{\beta}{i}_A \cvect{\beta}{j}_A } , \\
		(\cvect{u}{0})_B & = \sqrt{\metr_{00} (B) + 2 \metr_{0i} (B) \, \cvect{\beta}{i}_B + \metr_{ij} (B) \, \cvect{\beta}{i}_B \cvect{\beta}{j}_B } ,
	  \end{align}
	\end{subequations}
	with $\cvect{\beta}{i}_A$ and $\cvect{\beta}{i}_B$ the $i$-th component of the coordinate velocities of $\obs_A$ and $\obs_B$, respectively:
	\begin{equation}
	  \cvect{\beta}{i}_A = \frac{1}{c} \frac{\dd \cvect{x}{i}_A}{\dd t} , \qquad \cvect{\beta}{i}_B = \frac{1}{c} \frac{\dd \cvect{x}{i}_B}{\dd t} .
	\end{equation}
	
	\vspace*{0.3cm}
	\noindent {\large\bf 4.2 Quasi-Minkowskian path}
	\smallskip

\begin{figure}[t]
	  \begin{center}
		\includegraphics[scale=0.37]{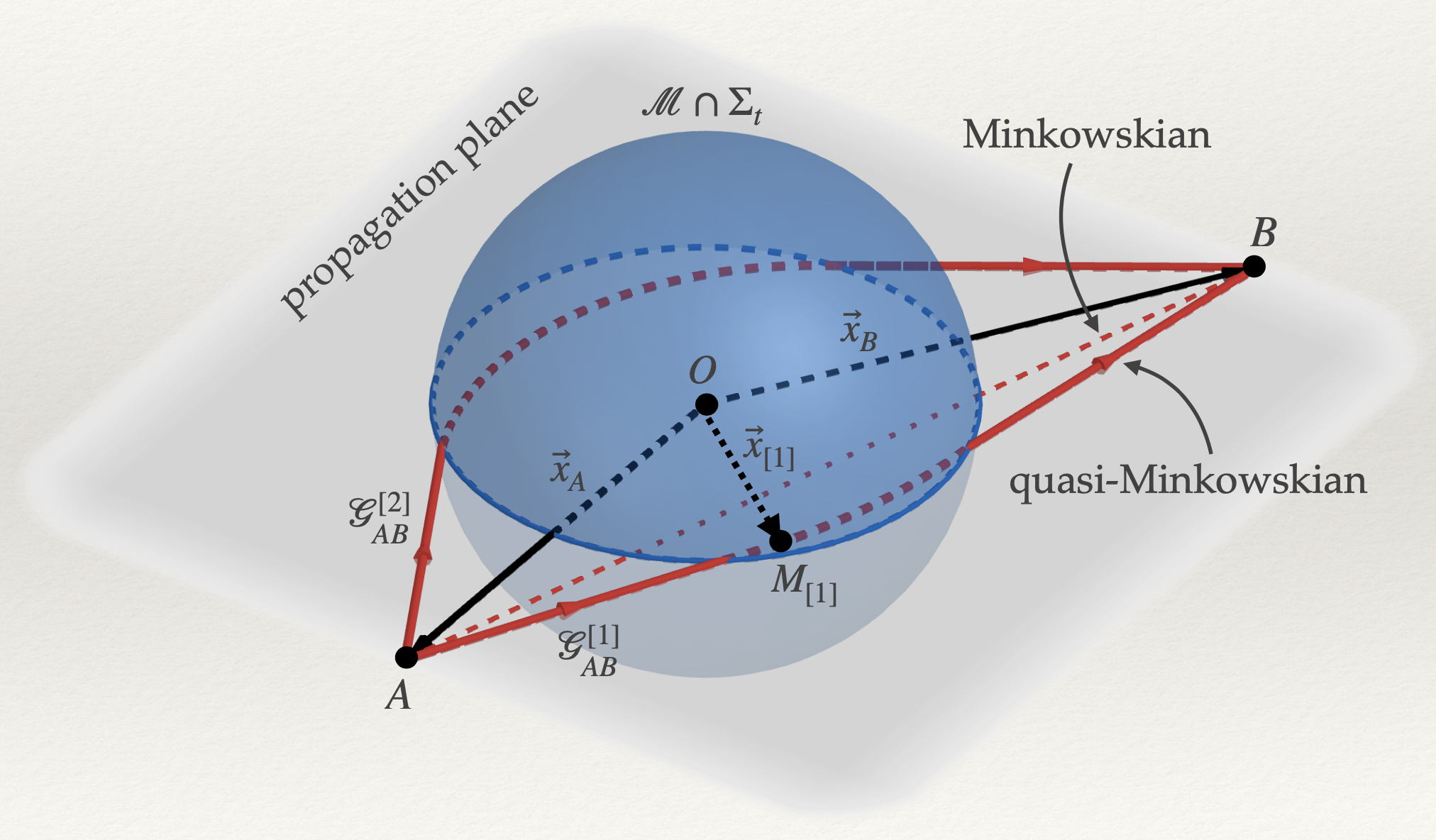}
		\caption{Representation of light rays $\lightray{AB}{1}$ and $\lightray{AB}{2}$ (\emph{in red}) between $A$ and $B$. $t_A^{[1]}$ and $t_A^{[2]}$ are the time at emission of the two rays. In this example, the emitter in $A$ and receiver in $B$ are on both sides of an optical medium with a spheric distribution (e.g., an atmosphere). The 3-volume occupied by the medium is $\partial \dommed = \dommed \cap \Sigma_t$, where $\Sigma_t$ is the hypersurface of simultaneity, at instant $t$, of an inertial observer $\obs$ at rest in the global coordinate system $( \cvect{x}{\alpha} )$. The light ray being the closer to the \emph{Minkowskian path} (\emph{in dash red}), that is to say the straight line between $A$ and $B$, is called a \emph{quasi-Minkowskian path} and can be determined iteratively.}
		\label{fig_03}
	  \end{center}
	\end{figure}

	In general, given the time of reception and the positions of both emitter and receiver, we cannot expect $t_A$ (and thus $\ttf$) to be unique (see Linet and Teyssandier 2016). Instead, we must admit that there exists a family of light rays $\lightray{AB}{\sigma}$ originating from $\vect{x}_A$ and received at $\vect{x}_B$ at instant of coordinate time $t_B$ (see example with $\sigma = \{ 1 , 2 \}$ in Fig. \ref{fig_03}). To each light ray we associate a unique time transfer function $\ttf^{[\sigma]}$ so that each time of emission is thus given by
	\begin{equation}
	  t_A^{[\sigma]} = t_B - \ttf^{[\sigma]} ( \vec{x}_A , t_B , \vec{x}_B ) .
	\end{equation}

	Each light ray $\lightray{AB}{\sigma}$ satisfies the eikonal equation \eqref{eq:eik} along $\lightray{AB}{\sigma}$. It can be evaluated at any point-event $M_{[\sigma]}$ of coordinates $\Qvect{x}_{[\sigma]} = (ct_{[\sigma]} , \vec{x}_{[\sigma]})$, it thus reads as follows:
	\begin{align}
	  & \bmetr^{00} \left( c t_B - c \ttf^{[\sigma]} ( \vec{x}_{[\sigma]} , t_B , \vec{x}_B ) , \vec{x}_{[\sigma]} \right) + 2 c \bmetr^{0i} \left( c t_B - c \ttf^{[\sigma]} ( \vec{x}_{[\sigma]} , t_B , \vec{x}_B ) , \vec{x}_{[\sigma]} \right) \frac{\partial \ttf^{[\sigma]}}{\partial x_{[\sigma]}^i} \nonumber \\
	  & \hspace{1.5cm} + c^2 \bmetr^{ij} \left( c t_B - c \ttf^{[\sigma]} ( \vec{x}_{[\sigma]} , t_B , \vec{x}_B ) , \vec{x}_{[\sigma]} \right) \frac{\partial \ttf^{[\sigma]}}{\partial x_{[\sigma]}^i} \frac{\partial \ttf^{[\sigma]}}{\partial x_{[\sigma]}^j} = 0 .
	\end{align}

	In most applications, the light path of interest is the quasi-Minkowskian path (see $\lightray{AB}{1}$ in Fig.~\ref{fig_03}) which is usually the trajectory with the minimal optical length. It can be looked for when the optical metric admits the following expansion (see Bourgoin 2020):
	\begin{equation}
	  \bmetr^{\mu\nu} ( P ; \varepsilon ) = \eta^{\mu\nu} + \sum_{\ell = 1}^{+\infty} \varepsilon^\ell \kappa^{\mu\nu}_{(\ell)} (P) , 
	\end{equation}
	with $\varepsilon$ a book-keeping parameter that we take to be small during our manipulations; at the end of our calculations we reset it to $\varepsilon = 1$. 
	
	For the rest of the paper, we shall assume that the quasi-Minkoskian path does exist and is actually unique. We thus intent to determine the time transfer function (which is also unique) within the quasi-Minkowskian path approximation. Therefore, it is natural to search for this time transfer function substituting the following ansatz into the eikonal equation:
	\begin{equation}
	  \ttf ( \vec{x}_A , t_B , \vec{x}_B ) = \frac{R_{AB}}{c} + \frac{1}{c}\sum_{\ell = 1}^{+\infty} \varepsilon^\ell \Delta_{(\ell)} ( \vec{x}_A , t_B , \vec{x}_B ) ,
	\end{equation}
	where $R_{AB} = \Vert \vec{x}_B - \vec{x}_A \Vert$.
	
	It is then possible to derive integro-differential equations for the delay function $\Delta_{(\ell)}$ in terms of metric components $\kappa^{\mu\nu}_{(\ell)}$---see Teyssandier and Le Poncin-Lafitte 2008 for light rays propagating in a vacuum and Bourgoin 2020 for extension to neutral optical media.

	\vspace*{0.7cm}
	\noindent {\large\bf 5. APPLICATION TO RADIO OCCULTATION EXPERIMENTS}
	\smallskip

	In this section, in order to highlight the capabilities of the formalism, we apply it to radio occultation by planetary atmospheres (see Bourgoin \emph{et al.}, 2021). We first restrict our attention to stationary optical spacetime omitting gravity terms before refractivity. A closed form solution to the time transfer function is found at first post-Minkowskian order for a spherically symmetric atmosphere in hydrostatic equilibrium. It is compared to numerical solutions to the canonical equations of relativistic geometrical optics.

	\newpage

	\noindent {\large\bf 5.1 Stationary spacetime}
	\smallskip

	For planetary atmospheres, whose neutral component may be seen as a tenuous non-dispersive optical medium, the refractivity $N(x)$ is usually a small quantity. To fix ideas, we can set $\forall \, P \in \dommed$, $N(P) \sim \varepsilon$ during our manipulations. We invoke an inertial observer $\obs$ of worldline $\wrdline$ inside $\dommed$ to which we attach the global coordinate system $(\cvect{x}{\alpha})$. Within $(\cvect{x}{\alpha})$, we suppose that the optical metric is stationary (i.e., $n( \vect{x} )$ and $\Qvect{w}( \vect{x} )$, that is to say $\partial_0 n = 0$ and $\partial_0 \cvect{w}{\mu} = 0$). Thus, $\cflin{k}{0}$ is a first integral along the light path (see time component of Eq. \eqref{eq:caneqsb}) meaning that the time transfer function is actually independent of $t_B$ (see Eq.~\eqref{eq:dirstatc}). In addition, we neglect gravity effects before refractivity, namely $U(\vect{x}) \ll \varepsilon c^2$ with $U( \vect{x} )$ the gravitational potential evaluated at position $\vect{x}$. With these conditions the physical spacetime metric reduces to $\metr_{\mu\nu} = \eta_{\mu\nu}$, which implies
	\begin{equation}
	  \bmetr^{\mu\nu} ( \vect{x} ; \varepsilon ) = \eta^{\mu\nu} + \varepsilon \kappa^{\mu\nu}_{(1)} ( \vect{x} ) + \varepsilon^2 \kappa^{\mu\nu}_{(2)} ( \vect{x} ) ,
	\end{equation}
	where the components $\kappa^{\mu\nu}_{(1)}$ and $\kappa^{\mu\nu}_{(2)}$ are given by
	\begin{equation}
	  \kappa^{\mu\nu}_{(1)} = 2 N w^\mu w^\nu , \qquad \kappa^{\mu\nu}_{(2)} = N^2 w^\mu w^\nu .
	\end{equation}

	In the following, we consider that the medium's rest frame is rotating with respect to $\obs$ at a constant rate $\vect{\omega}$. In the global coordinate system $(\cvect{x}{\alpha})$, the 4-velocity of the medium has components: $\forall \, \vect{x} \in \partial \dommed$ (see Fig. \ref{fig_03}), $\cvect{w}{0} ( \vect{x} ) = \Gamma ( \vect{x} )$ and $\cvect{w}{i} ( \vect{x} ) = \Gamma ( \vect{x} ) \, \cvect{\xi}{i} (\vect{x})$ with $\Gamma$ the Lorentz factor and $\vect{\xi}$ the coordinate velocity (divided by $c$) of the medium relative to $\obs$:
	\begin{equation}
	  \vect{\xi} ( \vect{x} ) = \frac{\vect{\omega} \times \vect{x}}{c} , \qquad \Gamma ( \vect{x} ) = \frac{1}{\sqrt{1 - \vect{\xi} ( \vect{x} ) \cdot \vect{\xi} ( \vect{x} ) }} .
	\end{equation}

\begin{figure}[t]
		\begin{center}
		  \includegraphics[scale=0.37]{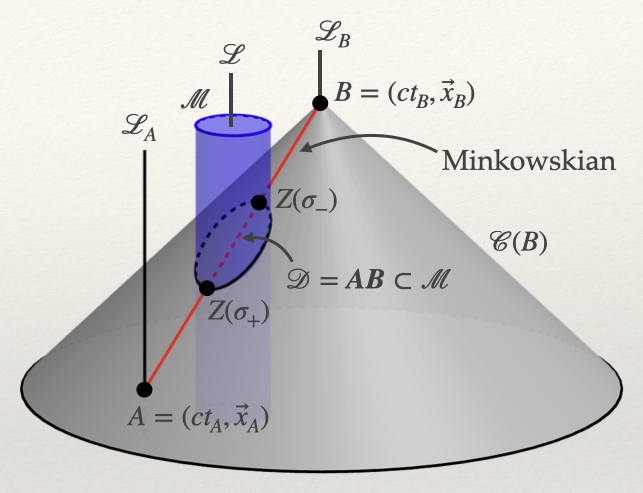}
		  \caption{Spacetime diagram representing a radio occultation experiment. The atmosphere draws a time-like tube $\dommed$ in spacetime (\emph{in blue}). The worldline of $\obs_A$ intersects the past-light cone $\cone(B)$ of the receiver $B$ at point-event $A$. The Minkowskian light ray (\emph{in red}) joining $A$ and $B$ intersects the optical medium $\dommed$ at two point-events $Z(\lambda_+)$ and $Z(\lambda_-)$. The domain of integration $\domint$ (\emph{in dashed red}) is restricted to the portion of the Minkowskian path $\boldsymbol{AB}$ that is included into $\dommed$.}
		  \label{fig_04}
		\end{center}
	  \end{figure}

	Then, the first two terms in the expansion of the delay function reads as follows
	\begin{subequations}
	  \begin{align}
		\Delta_{(1)} ( \vect{x}_A , \vect{x}_B ) & = \gamma^2 R_{AB} \int_{\domint} \left( \Gamma^2 N \right)_{\vect{z}(\lambda)} \dd \lambda , \\
		\Delta_{(2)} ( \vect{x}_A , \vect{x}_B ) & = \tfrac{1}{2} \gamma^2 R_{AB} \int_{\domint} \left( \Gamma^2 N^2 \right)_{\vect{z}(\lambda)} \dd \lambda + 2 \gamma R_{AB} \int_\domint \big( \Gamma^2 N \xi^i \big)_{\vect{z}(\lambda)} \bigg[ \frac{\partial \Delta_{(1)}}{\partial \cvect{x}{i}} \bigg]_{(\vect{z}(\lambda),\vect{x}_B)} \mathrm{d} \lambda \nonumber \\
		& - \tfrac{1}{2} R_{AB} \int_\domint \delta^{ij} \bigg[ \frac{\partial \Delta_{(1)}}{\partial \cvect{x}{i}} \frac{ \partial \Delta_{(1)} }{ \partial \cvect{x}{j} } \bigg]_{(\vect{z}(\lambda) , \vect{x}_B)} \mathrm{d} \lambda , \\
		\Delta_{(3)} ( \vect{x}_A , \vect{x}_B ) & = \ldots ,
	  \end{align}
	\end{subequations}
	where $\domint$ is the domain of integration (represented in Fig. \ref{fig_04}) and where $Z(\lambda)$ with $0 \leq \lambda \leq 1$ represents a point along the (zeroth-order) Minkowskian path between $A$ and $B$, that is to say $Z(\lambda) = ( ct_B - \lambda R_{AB} , \vect{z} ( \lambda ) )$ with $\vect{z} ( \lambda ) = \vect{x}_B -\lambda R_{AB} \vect{N}_{AB}$, $\vect{N}_{AB}$ being $(\vect{x}_B - \vect{x}_A) / R_{AB}$; we thus have $Z(0) = B$ and $Z(1) = A$. The parameter $\gamma$ is related to the \emph{light-dragging} coefficient $\delta_{\mathrm{drag}}$ by the following expression:
	\begin{equation}
	  \delta_{\mathrm{drag}} \equiv \gamma - 1 = \frac{ \vect{\omega} \cdot ( \vect{x}_B \times \vect{N}_{AB} ) }{ c } .
	\end{equation}

	  In most applications, $\gamma$ is arbitrarily fixed to $1$ forcing in this way the light-dragging contribution to vanish in the time transfer and then in the frequency transfer. The consequences of such a choice in modern experiments and observations will be further discussed in the conclusion.

	  \vspace*{0.3cm}
	  \noindent {\large\bf 5.2 Spherical symmetry}
	  \smallskip  

	  Let assume that $\dommed$ is filled with a \emph{perfect fluid} medium, rapidly rotating with respect to $\obs$ ($\omega a / c \sim 0.1$ with $a$ the equatorial radius of the occulting planet). In addition, assuming a spherically symmetric atmosphere, we have $N ( r )$ with $r = \Vert \vect{x} \Vert$. In these conditions, the refractivity profile can be modeled according to the hydrostatic equilibrium assumption, namely as a superposition of a polynomial function of degree $d$ (and coefficients $b_0 , \ldots , b_d$) describing the temperature profile and an exponential of scale heigh $H$ for the pressure profile:
	  \begin{equation}
		N ( r ) = N_a \exp \left( - \frac{r - a}{H} \right) \sum_{m=0}^{d} b_m r^m . 
		\label{eq:refr}
	  \end{equation}

	  Inserting this refractivity modeling into the expansion of the delay function, we can then determine the time transfer function (expression given at first order in $N_a$ and $\delta_{\mathrm{drag}}$) 
	  \begin{align}
		\ttf (\vec{x}_A , \vec{x}_B ) & = \frac{ R_{AB} }{c} + \frac{ \sqrt{ 2 \pi } H N_a }{ c } \big( 1 + 2 \delta_{\mathrm{drag}} \big) \exp \bigg( \! - \frac{ K - a }{ H } \bigg) \bigg. \nonumber \\
		& \times  \sum_{m=0}^{\infty} \frac{ ( 2m - 1 )!! }{ 2^m } \bigg( \frac{ H }{ K } \bigg)^{ \! m - 1/2 } \, \sum_{n=0}^{m_d} Q_{m-n} \, \sum_{l=n}^{d} \frac{ n! }{ l! ( n - l )! } b_l K^l ,
		\label{eq:ttf}
	  \end{align}
	  where $K ( \vect{x}_A , \vect{x}_B ) = \Vert \vec{N}_{AB} \times \vec{x}_B \Vert$. Numerical coefficients $Q_m$ and $m_d$ are defined in Bourgoin \emph{et al.}, 2021. This expression allows one to determine the time transfer and the frequency transfer (via Eqs. \eqref{eq:dirstat} and \eqref{eq:nuBsnuAli}) at linear order in the refractivity $N_a$. We can test the validity of this relation by direct comparison with solutions to a numerical ray-tracing from canonical equations~\eqref{eq:caneqs}. 

	  \vspace*{0.3cm}
	  \noindent {\large\bf 5.3 Numerical ray-tracing}
	  \smallskip

\begin{figure}[t]
		\begin{center}
		  \includegraphics[scale=1.]{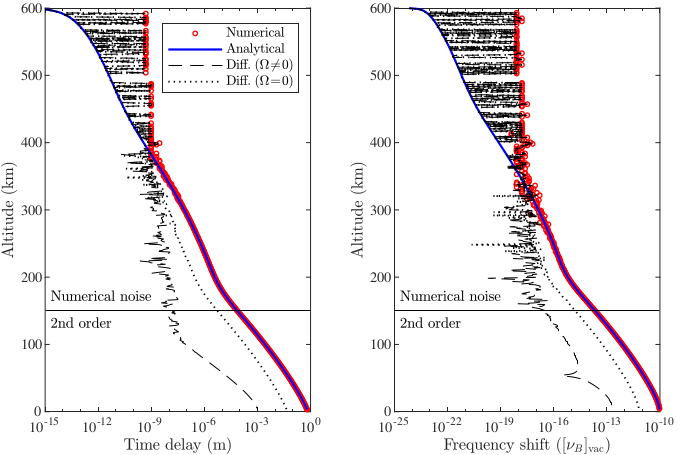}
		  \caption{Time delay (\emph{left panel}) and frequency shift (\emph{right panel}) due to a spherically symmetric atmosphere with $N_a=10^{-6}$. Solutions to a numerical integration of the canonical equations for relativistic geometric optics (\emph{red circles}) are compared with analytical predictions (\emph{blue curves}) from the time transfer function formalism. The differences between numerical and analytical results are represented \emph{in dash black} and \emph{in dot black}. The \emph{dashed black curve} corresponds to the difference considering the light-dragging effect, while the \emph{dotted black curve} represents the same difference neglecting it (i.e., considering $\delta_{\mathrm{drag}} = 0$ in the time transfer function).}
		  \label{fig_05}
		\end{center}
	  \end{figure}

	  Numerical integration of the canonical equations \eqref{eq:caneqs} through a planetary atmosphere can be achieved following part of the approach by Schinder \emph{et al.}, 2015. We assume an atmosphere with refractivity profile \eqref{eq:refr} and propagates numerically optical rays through it from an emitter $A$ (a spacecraft orbiting about the planet) to a receiver $B$ (at infinity). Because of refraction inside $\dommed$, the initial pointing $(\cflin{l}{i})_A$ must be iteratively corrected so that the optical ray eventually intercept $B$. From the determination of the wave-covector after crossing through the atmosphere, the time transfer and frequency transfer are eventually computed (see Eqs. \eqref{eq:defttf}--\eqref{eq:nuBsnuAli}) and then compared to their analytical counterparts deduced from \eqref{eq:ttf} (see Fig. \ref{fig_05}).

	  The difference (see \emph{dashed curves}) between numerical and analytical solutions is in perfect agreement (i.e., at the level of the numerical noise) above 150 km altitude. However, if the light-dragging effect is turned off in the analytical solution (see \emph{dotted curves}), the precision of the analytical solution decreases by two to three orders of magnitude at 150~km altitude. Below 150~km altitude, the precision of analytical solutions degrades gradually as the refractivity increases due to neglected second order terms in $N_a$ in \eqref{eq:ttf}.


	
	

	\vspace*{0.7cm}
	\noindent {\large\bf 6. CONCLUSION}
	\smallskip
	
    Gordon's optical metric and time transfer function formalism are two efficient theoretical tools for computing the time and frequency transfers in the presence of non-dispersive media. When applied to radio occultations experiment it demonstrated its efficiency (see Bourgoin \emph{et al.}, 2021). As a benefit, having an analytical description of the way time and frequency transfers change across the atmospheric profile allows one to easily assess the errors in the thermodynamic profiles (see Bourgoin \emph{et al.}, 2022). This last point is of a great interest in the context of radio occultations where the determination of error profiles usually require \emph{ad hoc} numerical procedures. In addition, the formalism presented in this paper, accurately considers the light-dragging effect which is at the threshold of being detected/needed in many experiments. As an example, let us mention that the light-dragging by the zonal winds in Saturn's atmosphere must be taken into account when processing radio occultations data of Cassini (see Schinder \emph{et al.}, 2015). In the context of the Earth, a direct application for this work is the modeling of the tropospheric delays while processing data of astro-geodetic techniques (SLR, LLR, GNSS, VLBI, etc.). Depending on the reference frame used for the data processing, BCRS ($\xi \sim 10^{-4}$) or GCRS ($\xi \sim 10^{-6}$), neglecting the light-dragging effect can induce errors on the time delay at the level of 0.05~mm and 3~mm in GCRS and BCRS, respectively. Concerning the Doppler, the error could reach the level of 0.01~$\mu$m/s and 10~$\mu$m/s in GCRS and BCRS, respectively. Comparing with current data precisions (see Sect. 1), it is evident that the light-dragging effect shall be modeled accurately in the next future.
	
	%
	%
	%
	%

	\vspace*{0.7cm}
	\noindent{\large\bf 7. REFERENCES}
	{
	
	\leftskip=5mm
	\parindent=-5mm
	\smallskip
	
    A. Bourgoin, 2020, ``General expansion of time transfer function in optical spacetime'', Phys. Rev. D 101, 064035.

	A. Bourgoin, M. Zannoni, L. Gomez Casajus, P. Tortora, and P. Teyssandier, 2021, ``Relativistic modeling of atmospheric occultations with time transfer funnctions'', \aa\ 648, A46.

    A. Bourgoin, E. Gramigna, M. Zannoni, L. Gomez Casajus, and P. Tortora, 2022, ``Determination of uncertainty profiles in neutral atmospheric properties measured by radio occultation experiments'', Adv. in Space Research 70, 8.

    C. Le Poncin-Lafitte, B. Linet, and P. Teyssandier, 2004, ``World function and time transfer: general post-Minkowskian expansions'', Class. Quantum Grav. 21, 4463-4483.

	B. Linet and P. Teyssandier, 2016, ``Time transfer functions in Schwarzschild-like metrics in the weak-field limit: A unified description of Shapiro and lensing effects'', Phys. Rev. D 93, 044028. 

	J. L. Synge, 1960, ``Relativity: The General Theory'', North-Holland Publishing Company, Amsterdam.

	P. J. Schinder, F. M. Flasar, E. A. Mourad, R. G. French, A. Anabtawi, E. Barbinis, A. J. Kliore, 2015, ``A numerical technique for two-way radio-occultations by oblate axisymmetric atmospheres with zonal winds'', Radio Science 50, 712-727.

	P. Teyssandier and C. Le Poncin-Lafitte, 2008, ``General post-Minkowskian expansion of time transfer functions'', Class. Quantum Grav. 25, 145020. 
	
	
	
	}
	
	\end{document}